\RequirePackage{fix-cm}
\documentclass[smallextended]{svjour3}       % onecolumn (second format)
\smartqed  % flush right qed marks, e.g. at end of proof
\usepackage{graphicx}
\usepackage{natbib,epsfig,chngcntr}
\usepackage{amsmath,amssymb}
\newcommand{\sun}{\odot}
\newcommand{\apj}{ApJ}
\newcommand{\apjl}{ApJ}
\newcommand{\aj}{AJ}
\newcommand{\araa}{ARA\&A}
\newcommand{\mnras}{MNRAS}
\newcommand{\aap}{A\&A}
\newcommand{\etal}{{et.\,al.}}
\begin{document}
\title{Gamma-ray Bursts and the Early Star-formation History%\thanks{Grants or other notes
}
\subtitle{GRBs and $z>4$ Star-formation}


\author{R. Chary, P. Petitjean, B. Robertson, M. Trenti, E. Vangioni}


\institute{R. Chary \at
                     U.S. Planck Data Center, Caltech, MS314-6, Pasadena, CA91125, USA
                            \email{rchary@caltech.edu}           %  \\
           \and
           P. Petitjean \at
           	Institut d'Astrophysique de Paris, France \email{petitjean@iap.fr}
	 \and
	 B. Robertson \at
	 University of California, Santa Cruz, CA, USA \email{brant@ucsc.edu}
	 \and
	 M. Trenti \at
	 University of Melbourne, Victoria, Australia \email{mtrenti@unimelb.edu.au}
	 \and
	 E. Vangioni \at
	 Institut d'Astrophysique de Paris, France \email{vangioni@iap.fr}
           }

\date{Received: date / Accepted: date}

\maketitle

\begin{abstract}
We review the uncertainties in high-z star-formation rate (SFR) measures and the 
constraints that one obtains from high-z gamma-ray burst (GRB) rates on them. We show that
at the present time, the GRB rates per unit star-formation at $z>3$ are higher
than at lower redshift. There could be a multitude of reasons for this: a stellar metallicity
bias for GRB production, a top-heavy initial mass function (IMF) and/or missing a significant
fraction of star-formation in field galaxy surveys due to incompleteness, surface brightness limitations and cosmic variance. 
We also compare metallicity predictions made using a hierarchical model of cosmic chemical 
evolution based on two recently proposed SFRs,  one based on the observed
galaxy luminosity function at high redshift and one based on the GRB rate and find
that within the considerable scatter in metal abundance measures, they both are consistent with the data. 
Analyzing the ensemble of different measurements together, 
we conclude that despite metallicity biases, GRBs may be a less biased probe of star-formation at $z>3$ than at $z<2$.
There is likely to be a common origin to the high GRB rate per unit star-formation and the high observed Lyman-continuum production
rate in high redshift galaxies and that this may be due to a relative overabundance of stars with mass $>$25\,M$_{\sun}$ which are likely GRB progenitors.
We also find that to reconcile these measurements with the Thomson scattering cross section of cosmic microwave background (CMB) photons measured
by Planck,
the escape fraction of Lyman-continuum photons from galaxies must be low, about $\sim$15\% or less and that the clumping factor of the IGM is likely to be small, $\sim$3.
Finally, we demonstrate that GRBs are unique probes of metallicity evolution in low-mass galaxy samples
 and that GRB hosts likely lost a significant fraction of metals to the intergalactic medium (IGM) due to feedback processes such as stellar winds and supernovae.

\keywords{Gamma ray bursts \and Star formation history \and Reionization \and Chemical abundances}
\end{abstract}

\section{Introduction}
Measuring the evolution of the co-moving star-formation rate density (SFRD) with redshift has been
one of the major goals of galaxy evolution studies. Primary tracers of star-formation
include the rest-frame ultraviolet continuum, the mid-infrared through sub-millimeter continuum
from thermal dust emission, 1.4 GHz and 30 GHz radio emission which traces the synchrotron
and free-free components respectively and nebular lines (e.g. H$\alpha$ and H$\beta$). 
Each of these tracers have different systematics associated with them. Ultraviolet wavelengths
are affected by dust attenuation and also require uncertain extrapolations for the faint end
of the galaxy luminosity function. Mid-infrared and sub-millimeter tracers are biased towards
the bright end of the galaxy luminosity function and suffer from AGN contamination. Current
radio surveys are insensitive to all but extreme, ultra-luminous galaxies. Nebular lines are powerful
but require extensive spectroscopic follow-up which is limited to the bright end of the LF. Furthermore,
they too suffer from attenuation and stellar absorption which needs to be corrected for.

The goal has motivated a multitude
of deep surveys; they have been undertaken at visible/ultraviolet (UV) wavelengths with {\it Hubble}, in the
mid- and far-infrared with {\it ISO}, {\it Spitzer}, {\it Herschel} and more recently
{\it ALMA} and also with narrow-band filters and grisms (e.g. {\it Hubble}/WFC3 and NICMOS) to detect strong nebular
emission lines. These surveys have resulted in a generally broad agreement whereby the SFR
rises from redshift 0, peaks at $z\sim1-2$ and declines with increasing redshift out to $z\sim10$.
Our understanding of this has been based on observations of 
the bright end of the luminosity function with significant assumptions needed to extrapolate
to the faint end and to derive a co-moving star-formation rate density. For example, the UV SFR measurements
are sensitive down to few times 10$^{8}$\,L$_{\sun}$ while the mid- and far-infrared surveys only detect
galaxies down to 10$^{11}$\,L$_{\sun}$. As a result, it is unclear how much attenuation correction, if any, is 
required at the faint end of the galaxy LF and whether even the slope of the faint-end of the galaxy LF is robustly
constrained.

This is particularly true at the epoch of reionization at $z>6$, where considerable uncertainties exist at both the
bright end of the ultraviolet luminosity function and the faint end. Although 1000s of candidate galaxies have been detected out to $z\sim10$, 
these have been through deep, pencil-beam surveys or through observations of galaxies lensed by foreground clusters, both of which trace
relatively small comoving volumes at the redshifts of interest.
Since these surveys are done with 4\,arcmin$^{2}$ instrumental fields of view, there is the issue of cosmic variance. A comparison with semi-analytical models 
indicate that similar fields of view could show as much as a factor of 3 scatter in luminosity density at $z>6$, depending on the large scale structure.

Long-duration GRBs\footnote{Throughout this paper, we refer to long-duration GRBs whose gamma-ray emission typically lasts $>$2s and has a soft spectrum with hardness ratios of $\sim$0.5 compared to the short GRBs which last $<$2s, have hardness ratios of $\sim$1.0 and that
are thought to arise from merging double-degenerate systems.}, since they are thought to be the evolutionary end-states of massive stars,
 are an alternate technique for tracing the comoving SFRD \citep{Wijers, Blain2000, bromm2002a,daigne06b, ishida11}.  
 Provided the selection effects
associated with GRBs can be understood, measuring the co-moving rate density of GRBs and/or
comparing the properties of their host galaxies with field galaxies populations can allow
a calibration of GRBs as a star-formation rate metric. With rates of $\sim1$/day, they are quite abundant with
the biggest limitation arising in identification and spectroscopic follow-up of the afterglow \citep[e.g][]{Kruhler}. In addition, since GRB afterglows are bright, 
the spectroscopic follow-up detects absorption by metal lines that arise both in the host
galaxy and in the intervening IGM. This would hopefully allow the characterization of star-forming environments in distant galaxies, and their evolution with redshift, which is otherwise a challenge.
 
However, the question remains whether GRBs are unbiased tracers of star-formation. 
Are there metallicity/luminosity biases in the kinds of galaxies that GRBs occur in? Are
there environment biases in the sense that GRBs only occur in stars with large angular momentum \citep{WoosleyHeger}? 
Although accurate answers to these questions are unclear at the
present time, in this article, we
review the current state of using GRB rates as a star-formation rate metric and outline
possible ways forward.

\section{GRB Rates as a Tracer of Star Formation}
Long-duration gamma-ray bursts are thought to arise from the collapse of the core of a massive star through a black hole-accretion disk system. This knowledge has come about
due to the association between GRBs and star-forming galaxies \citep[e.g][]{Bloom, LeFloch, Chary2002} and the detection of a late-time core-collapse supernova light curve underlying many GRB afterglows \citep{Woosley}.
Since the lifetime of a massive
star is relatively short $\lesssim$10\,Myr relative to typical timescales for star-formation in a galaxy, measuring the rates of GRBs can in principle be a tracer of high-mass star-formation. 
If the stellar initial mass function is known, a significant source of uncertainty, they could even provide a global tracer of the star-formation rate density.

The primary disadvantage of using GRBs is there is no
mechanism to calibrate the mass range of the progenitor star and its physical properties such as metallicity, angular momentum and binarity to the GRB. For example, do high metallicity
stars have large enough opacity to radiation pressure and blow off most of their envelope through stellar winds inhibiting core-collapse? Do stars with large angular momentum lose too much
of their mass through magnetic torques? Models of stellar evolution have provided the first crucial insights into this process \citep{Heger}, yet observational data is limited.
Some theoretical attempts have been made to quantify the efficiency
of GRB production depending on metallicity and rotation of  
  progenitors \citep[see e.g][]{yoon06}, but a systematic and
  comprehensive study is still missing.

In the absence of such a physically motivated relationship between the formation of a massive star and the occurrence of 
a gamma-ray burst, one has to rely on empirical relationships between the GRB-rate
and star-formation rate while factoring in observational selection effects such as the fraction of GRBs that are detected above a particular gamma-ray fluence limit, the efficiency of spectroscopic
follow-up of the optical/near-infrared afterglow and the nature of galaxies in which the bulk of star-formation is taking place \citep{Vergani2015}. 

\subsection{Uncertainties in High Redshift Star Formation}

The high redshift ($z>3$) star-formation rate has, almost exclusively, been measured by using the rest-frame ultraviolet light from Lyman-break galaxies detected in deep optical surveys.
This has been combined with corrections for dust attenuation derived from the slope of the ultraviolet continuum to yield comoving star-formation rate densities. These measurements,
using ground-based (e.g. {\it Keck}/LRIS and {\it Subaru}/SuprimeCam) and {\it Hubble} deep surveys have revealed that the extinction corrected
star-formation rate density appears to monotonically decline as $\sim(1+z)^{-3.75}$ with increasing redshift between $3<z<8$ \citep{O14}. There are even hints of a steeper decline, $\sim(1+z)^{-10.5}$ at $8<z<10$
down to the observed luminosity limit \citep{Bouwens2015, O14}\footnote{These exponents are just parametric representations and are currently, not physically motivated.}.

There are however significant uncertainties associated with this measurement. Primary among these are:
\begin{itemize}
\item the contribution from galaxies at the faint end of the UV-luminosity function since the surveys typically reach depths of 0.04L$_{*,{\rm UV}}$ \citep{Trenti, Robertson2013};
\item dust obscuration which although constrained by the UV slope, are uncertain, due to the significant scatter in the relationship between UV slope and attenuation correction \citep{Battisti};
\item surface brightness limitations, with individual high-z galaxies appearing to be small, typically $\lesssim$1-2 kpc \citep{Oesch2010} which suggests that current surveys may be picking up the brightest star-forming knots
and missing extended, low surface-brightness star-formation;
\item cosmic variance since the amount of area covered is much smaller than the correlation length of large scale structure \citep{Trenti2008, Robertson2010};
\item stellar initial mass function which affects the translation from ultraviolet luminosity density to a star-formation rate density \citep{Chary2008}.
\end{itemize}
To illustrate, \citet{Kistler13} showed
that integrating the ultraviolet luminosity function down to an absolute magnitude $M_{\rm UV} = -10$ (significantly below the lowest
measured absolute magnitude value of $M_{\rm UV} \simeq -18$) would lead to an order of magnitude increase in the estimated SFR
at redshifts $z \sim 4$. However, it is unclear what the lower luminosity limit is for the integral over the luminosity function.
This is because the luminosity of the lowest mass dark matter halo which can sustain star-formation is uncertain due to the inefficiency
 of gas cooling in halos lower than mass $\sim$10$^{7}$\,M$_{\sun}$.
Since sub-L$_{*,{\rm UV}}$ galaxies may contribute the bulk of the SFRD, measurement of the faint- end slope of the UV luminosity function, where completeness corrections and surface brightness dimming issues are significant, needs to be undertaken carefully (Steidel et al. 1999; Bouwens et al. 2006). This uncertainty is further manifested in the change in derived luminosity function parameters obtained
 from  data presented in \citet{Bouwens2006} and \citet{Bouwens2015} where deeper data seem to show significant scatter in luminosity function parameters, even at the bright end.
 
  Similarly, if dust extinction were a significant issue, the galaxies that dominate the SFRD would be UV-faint or undetected in magnitude-limited rest-frame UV surveys. 
  Since dust can be produced efficiently, within a few years, in the shells of supernovae \citep[e.g.][]{Matsuura, Marassi}, the fraction of dust obscured star-formation
could be quite significant even at high redshift.
  Far-infrared/submillimeter observations would be required to identify these galaxies and measure their contribution to the bolometric luminosity density. Millimeter observations with ALMA, ACT and SPT are indeed finding significant populations of lensed, dust obscured galaxies at $z\sim5-6$, yet these are rare enough that their contribution to
 the bolometric luminosity density appears to be small, at the present time \citep{Weiss, Su}. Deep, forthcoming ALMA observations will help clarify the uncertainties due to dust obscuration at least
 in the limited areas of sky that can be observed with the small ALMA primary beam.
 
 At $z>6$, these uncertainties only get enlarged further. Although the UV measured SFR from Lyman-break galaxies,
the optical depth to electron scattering ($\tau$) measured for the CMB by {\it WMAP} and {\it Planck}, and other ancillary probes of the IGM neutrality at $z>6$ are arguably consistent with each other, each
of these has very significant uncertainties \citep[e.g.][]{Robertson2015}. For example, the measure of $\tau$ from {\it Planck} is 0.058$\pm$0.012. Since almost 80\% of the
optical depth ($\sim$0.04) arises from the ionized IGM at $z<6$,  it is merely a $\sim$1$\sigma$ constraint on
the ionizing photon production rate at $z>6$. Furthermore, translating the star-formation rate to a measure of the $\tau$
requires knowing the highly uncertain escape fraction of ionizing photons. While \citet{Robertson2015} assumed a value of 0.2 which is high, recent constraints suggest it may be
lower than 0.02 \citep{Grazian} resulting in a lower $\tau$ for a particular star-formation history. 

GRBs are relatively insensitive to these limitations. If the GRB rate density were correlated with the comoving SFRD at lower redshifts, where the 
UV and IR have been brought into agreement \citep[e.g][]{Reddy}, measurement of the GRB rate density at $z > 3 $ could provide an independent pathway to quantifying the SFRD (see also, e.g., Price et al. 2006).
Complementary measures of star-formation using probes such as long duration GRBs could thereby help address these uncertainties that persist.

\subsection{Implications from Gamma-Ray Burst Rates}
Using a small sample of $\sim$50 spectroscopically-confirmed GRBs over the first two years of the {\it Swift} mission, \citet{chary2007a} found that the long duration
GRB rate was constant over $0.5<z<3$ at a rate of 3.7$\pm$1.1$\times10^{-11}$\,Mpc$^{-3}$\,yr$^{-1}$. This number is undoubtedly an underestimate of the true rate obtained
by integrating over a GRB luminosity function, since it requires factoring in
the efficiency of optical afterglow detection and spectroscopic confirmation \citep[see e.g][]{Sun2015}. However, as long as the follow-up efficiency is constant over this redshift range, it can be considered a reasonable
lower limit. After cross calibrating against the SFR at $z<3$, the measured GRB rate can yield
a measure of the SFR at $z>3$, assuming a constant ratio between the GRB rate and star-formation rate. 
Although the uncertainties
are large, due to the small sample of high-z GRBs, the amount of SFR was a factor of $\sim$2-3 higher than what had been measured through field galaxies at $z\sim6$. 

This calculation however did not take into account evolution of the GRB luminosity function which more recent work has investigated \citep{Lien2014, Salvaterra2012, Qin2010}. This is equivalent to
factoring in some of the GRB production efficiencies discussed in Section 2. For example,
by taking a sample of spectroscopically confirmed {\it Swift} GRBs, \citet{Salvaterra2012} showed that there is evidence for a $\sim(1+z)^{2}$ luminosity or density evolution
with redshift in the GRB rates. Similarly, \citet{Qin2010} argued that the measured GRB rates require possibly both a metallicity threshold for GRB production and a redshift evolution with respect to the
star-formation rate density. Regardless, the conclusion that the GRB rate per unit star-formation increases with increasing redshift appears to be robust although the magnitude of enhancement
suffers from uncertainties in both the GRB rates and the SFRD.

More recently, \citet{Lien2014} have simulated the trigger algorithm adopted by the {\it SWIFT}/BAT instrument.
They compared the simulated peak flux and redshift distribution of GRBs 
with those observed. Their sample of 66 bursts only included bursts whose
redshifts arise from afterglow measurements rather than host galaxy measurements and thereby rejected a number of low-z GRBs.
They however obtained an order of magnitude higher GRB rate (0.4 Gpc$^{-3}$\,yr$^{-1}$) even at $z\sim0$.  The origin of this difference is far from clear and is most likely folded into the detection efficiency
as a function of peak energy of {\it Swift}. Regardless of the absolute comoving rate of GRBs, they too arrive at the same general conclusion that
the GRB rate per unit star-formation increases with increasing redshift \citep[see also][]{Kistler2008, Virgili2011}.

The presence of a significant amount of star formation below
  typical Lyman break galaxy survey limits at high $z$ can be probed
  independently through observations of GRB locations at rest-frame UV
  wavelengths ($\sim 0.15~\mathrm{\mu m}$) once the afterglow has
  faded. In fact, the fraction of undetected hosts provides a direct
  estimate of the amount of missed star formation to a given
  luminosity limit \citep{trenti12}. Based on ultradeep searches for a
  small sample of GRB hosts at $z>5$, which only yield upper limits,
  it is possible to infer that the galaxy luminosity function must
  extend with a steep faint-end slope ($\alpha \sim -1.8$) to at least
  $M_{AB}=-15$, about two mag fainter than the Hubble UDF
  observational limit for direct detection of galaxies
  \citep{trenti12,tanvir12}. One caveat is the small sample size (six
  targets) of the study. Indeed, more recent observations yielded the
  first rest-frame UV detections of GRB hosts at $z>5$
  \citep{mcguire16}, so future data on GRB hosts might lead to a
  revision of the fraction of missing star formation from Lyman break
  galaxy surveys.

\begin{figure}[h]
\center
\includegraphics[scale=0.3]{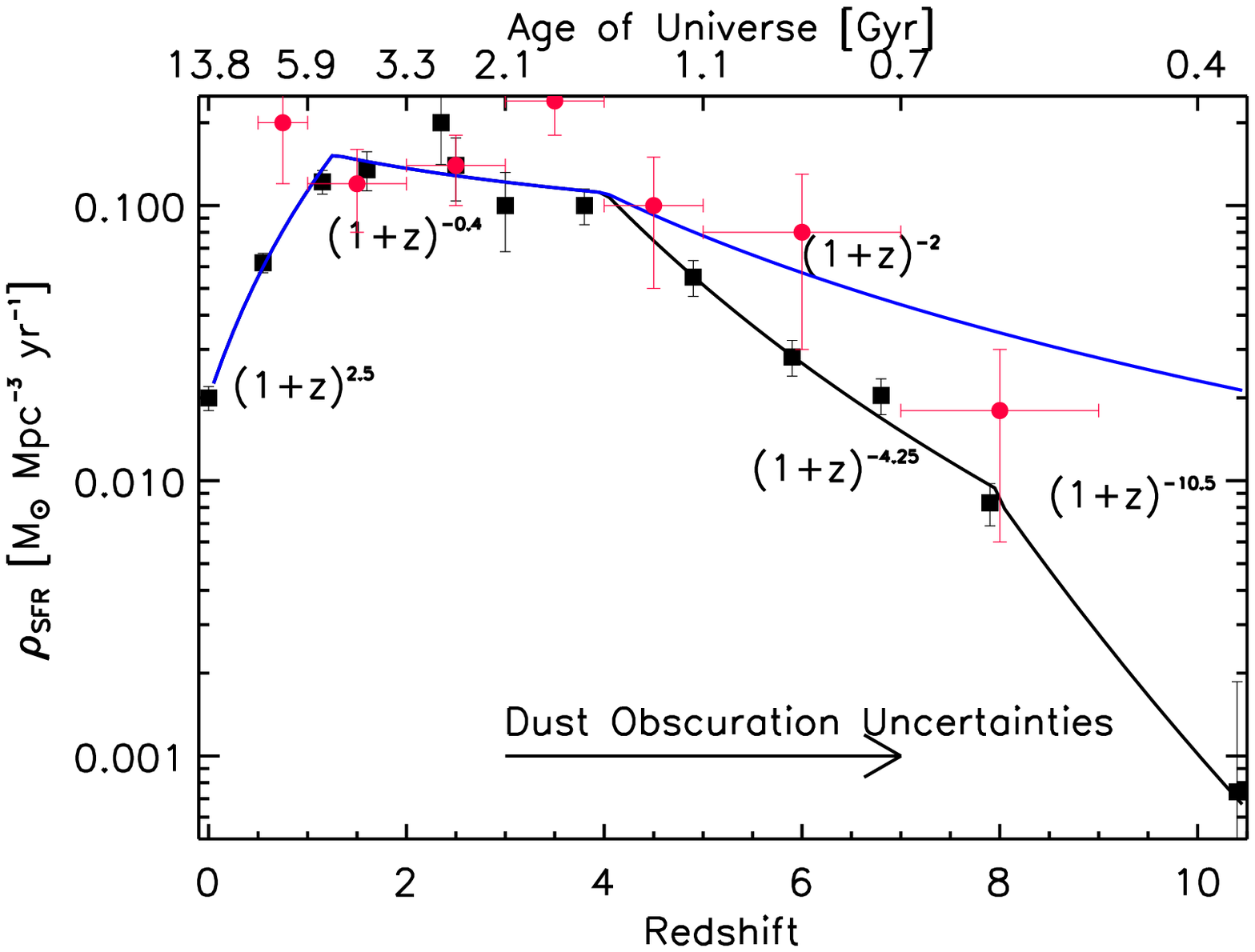}
\includegraphics[scale=0.3]{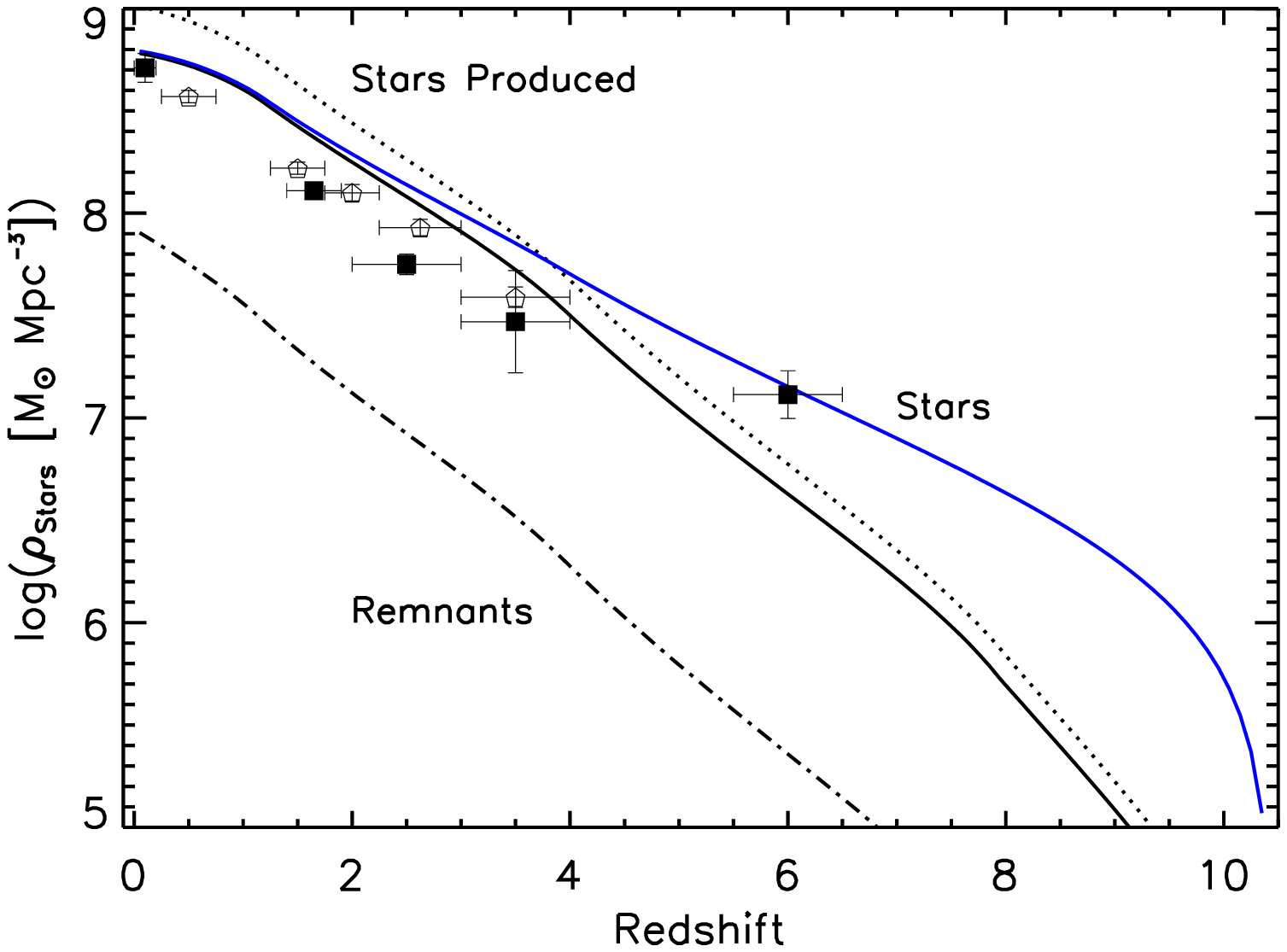}
\includegraphics[scale=0.3]{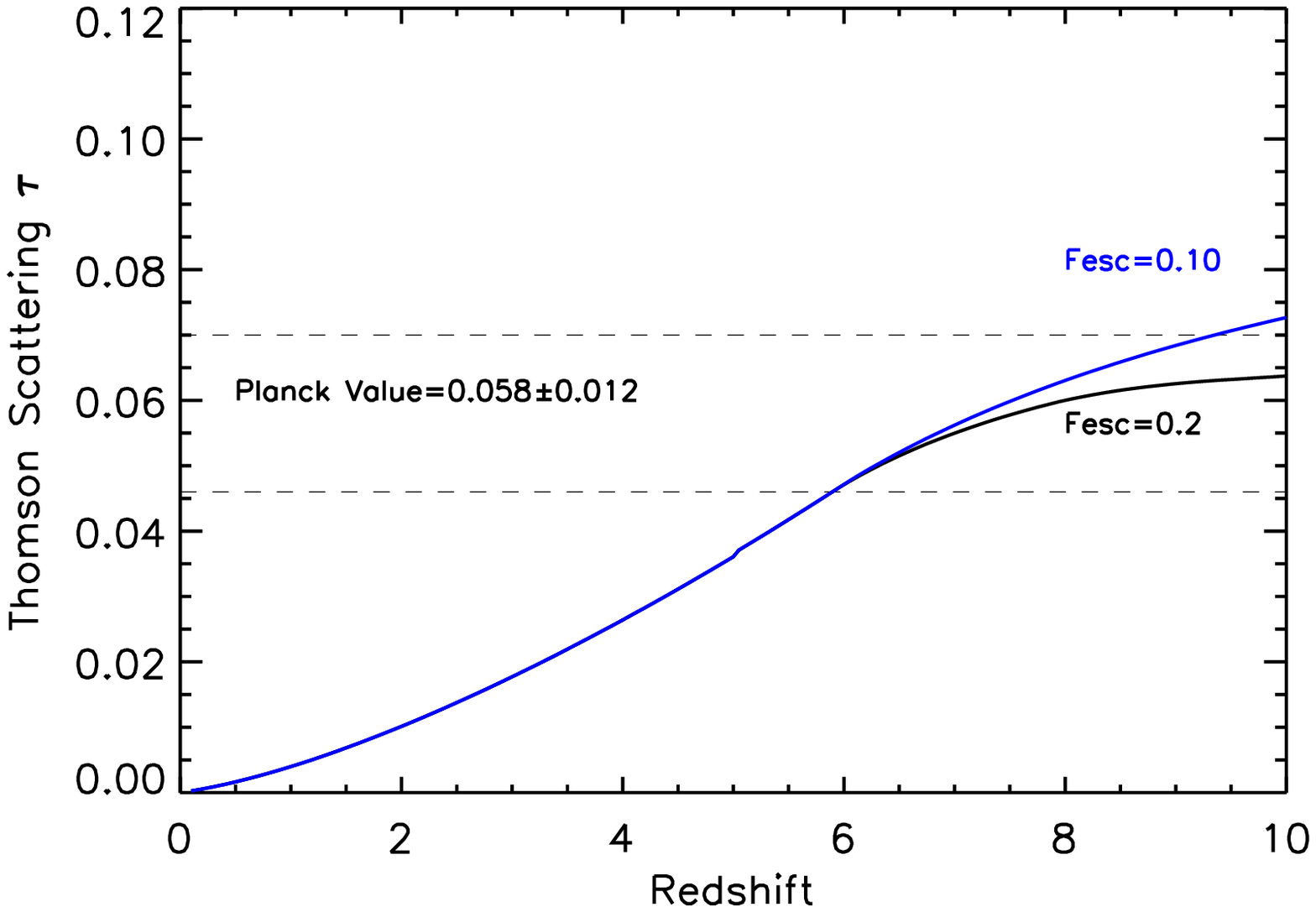}
\includegraphics[scale=0.3]{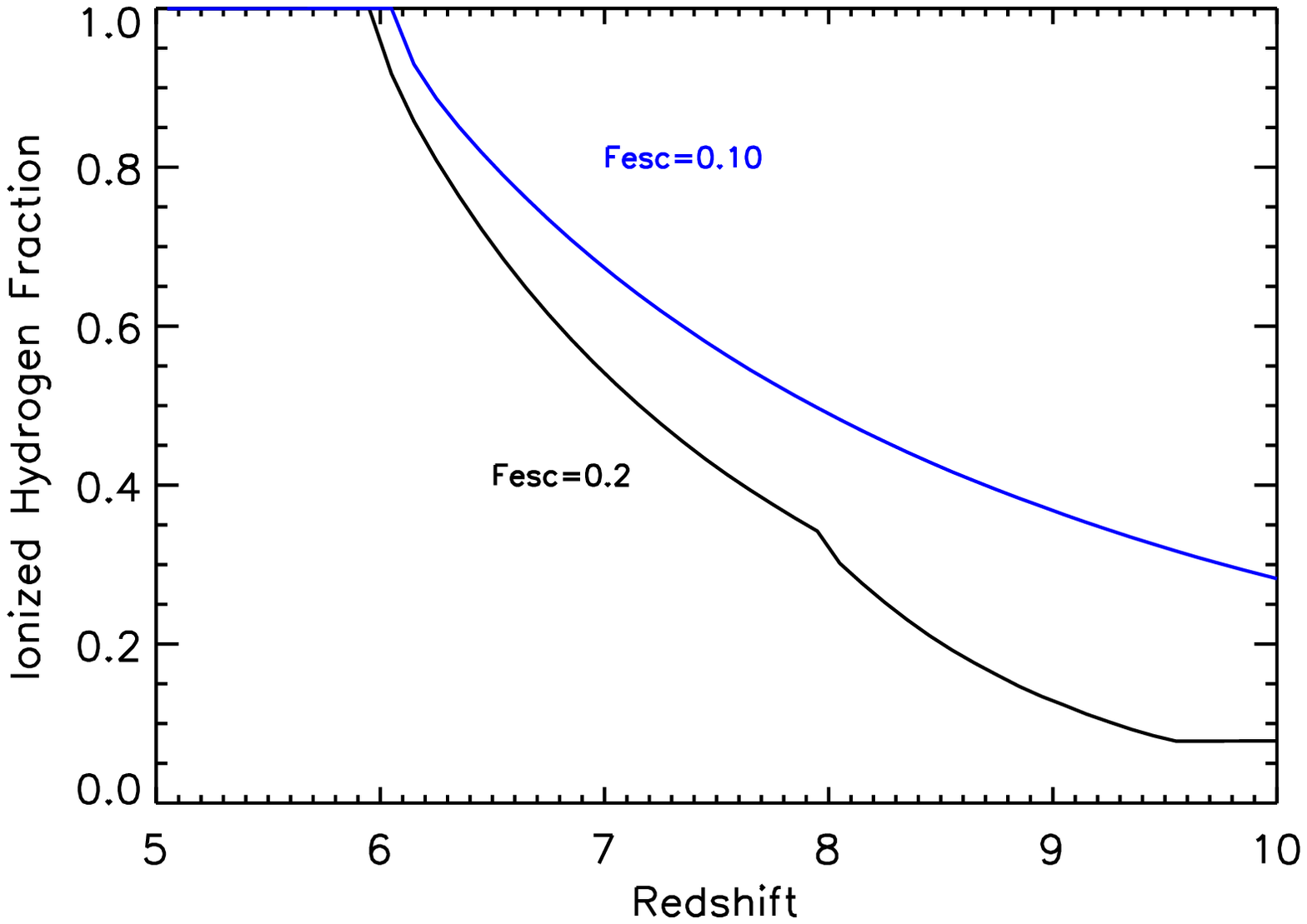}
\caption{The top left panel shows the comparison between the scaled GRB rate (red colored points) and the attenuation-corrected
star-formation rate density from deep fields (black points) illustrating that the GRB rate per unit star-formation increases with increasing redshift and at $z>4$ tends to lie above that inferred from the rest-frame ultraviolet observations. The redshift parameterizations of the black and blue lines
are shown with the blue line an elevated star-formation history that matches the GRB rates better. Black points are from \citep{Murphy2011, Reddy2009, Bouwens2015} while red points are from \citet{chary2007a}. 
The top right panel shows a comparison between the integral of the star-formation rate density from field galaxies (black line) and that from GRBs (blue line) with the stellar mass density inferred from rest-frame
visible light observations, assuming a standard Salpeter IMF. $z<4$ stellar mass densities are estimates from \citet{Marchesini} and \citet{Elsner} and at $z\sim6$ from \citet{Chary2008}.
The bottom left panel illustrates the Thomson scattering cross section derived for
CMB photons from {\it Planck} and that inferred from the SFRs displayed in panel 1, assuming an escape fraction of ionizing photons of 10-20\% and assuming the higher Lyman-continuum
production rate inferred in \citet{Shim2011}. 
The bottom right panel compares the ionized fraction of the IGM inferred from star-forming galaxies (black line) with that inferred from the GRB rate (blue line) with the biggest uncertainty in this translation arising from the unknown clumping factor of the IGM; here we adopt a value of 3 for the clumping factor. In this multi-panel figure, we highlight the implications of GRB-derived measures of
high-z star-formation and other measures of
star-formation. The current data favor a higher far-ultraviolet luminosity density at $z>4$ than inferred from field galaxy surveys, a high Lyman continuum production rate probably due to a top-heavy IMF, an escape fraction of $\sim$10-15\% and a lower clumping factor for the IGM than current simulations indicate \citep{Trac}.}
\label{sfropt}
\end{figure}

\citet{Robertson2012} used a larger sample of 112 from a {\it Swift} luminosity-limited
catalog to argue that relative to the star-formation rate, there seemed to be evidence for a slight evolution for the GRB rate of $(1+z)^{0.5}$ although the exponent strongly depends on the rate of optically-faint or ``dark" GRBs
as a function of redshift (Figure \ref{sfropt}). 
Since GRB hosts appear to be sub-luminous, low-metallicity galaxies \citep[e.g][]{Modjaz}, that would imply a large population of faint (sub-L$_{*}$) Lyman-break galaxies below the detection
limit of field galaxy surveys. \citet{chary2007a} estimated that this would imply a slope of the UV luminosity function of galaxies to be $\alpha=-1.9$ at $z\sim6$, compared to the measured
value of $-1.73\pm0.21$ derived by \citet{Bouwens2006}. Although the field galaxy slope was revised to $-1.87\pm$0.1 in \citet{Bouwens2015}, corresponding changes to the Schechter
luminosity function fit parameters ($\phi_{*}$ and M$_{*}$, the number density and characteristic luminosity respectively), have not significantly modified the total UV luminosity density and the discrepancy between the scaled GRB rate and the Lyman-break galaxy luminosity density remains a factor of $\sim$2. Unlike the redshift evolution for the SFRD quoted earlier, the SFR inferred from GRBs
would imply a much slower fall-off as a function of 
redshift, $\propto (1+z)^{-3}$ for $z> 4$ \citep{Kistler13,wang} or $\propto (1+z)^{-2}$ as shown in Figure 1. 

\citet{Robertson2012} also argued that the high SFRD inferred from GRBs would violate the stellar mass density at $z\sim4-6$ for canonical shapes of the IMF and
produce enough ionizing photons to result in a $\tau$ of 0.09 compared to the {\it Planck} derived value of 0.058$\pm$0.012. The former can be 
easily reconciled if the stellar IMF is more top-heavy (dn/dM$\propto$M$^{-1.7\pm0.1}$) at high redshift,  as seems to be indicated by nebular line observation of some $z>4$ galaxies \citep{Shim2011}
and has been previously argued at $z\sim6$ by \citet{Chary2008}. Such an IMF has a factor of 5 more massive stars above 25\,M$_{\sun}$ and would produce a factor of 5.4 more Lyman-continuum ionizing photons
per second but only a factor of 3.4 more far-ultraviolet photons at $1500$\AA. The star-formation rate in M$_{\sun}$\,yr$^{-1}$ corresponding to 
a particular measured far-ultraviolet luminosity density is reduced by the latter factor. 
Such an IMF would imply that a large fraction of the sources of ionizing photons quickly evolve into stellar remnants like
neutron stars and black holes, which would not be contributing to the rest-frame visible light luminosity density of galaxies at $z\sim6$ from which stellar mass densities are derived (Figure 1). 
Since GRBs are thought to arise from $>$25\,M$_{\sun}$ stars, 
such an IMF would imply a 1.6 times higher GRB rate per unit star-formation 
and thus be a convenient solution to the observed discrepancy between the SFRD observed from the UV light of galaxies and the GRB rate at high-z. Intriguingly it also reconciles
the discrepancy with the $z\sim6$ stellar mass density as described in \citep{Chary2008} and matches the slope of the extragalactic transient luminosity function at high-z seen by \citet{Sun2015}.

The discrepancy with the Thomson scattering optical depth can also be addressed. First, if the Lyman continuum production rate is as high as inferred in \citet[][assuming the H$\alpha$ to FUV continuum ratio is 0.08 rather than the 0.008 seen in the local Universe]{Shim2011} from observations
of H$\alpha$ in field galaxies, an escape fraction of 10-15\% would reduce the $\tau$ corresponding
to a particular star-formation history to be within {\it Planck} derived values (Figure 1). Second, as explained earlier, $\tau$ is only a weak constraint on star-formation at $z>6$. 
With such a weak $\sim1\sigma$ constraint, the current discrepancy with Planck Collaboration (2015) is hardly significant.
Reducing the uncertainty in the Thomson scattering optical depth by a factor of two with improved calibration of the {\it Planck} polarization data on large angular scales will help assess if
there is a tension with CMB polarization measurements.

Two other possibilities exist to explain the high GRB rate at $z>6$. One scenario would argue for a population of dust-obscured galaxies that have thus far not been detected. This argument
is disfavored
since dusty infrared luminous galaxies are high mass, high metallicity systems at least at $z<3$ where they have been extensively studied and dominate the star-formation rate density. 
At these low redshifts, GRB hosts do not sample the infrared luminous galaxy population, even including the significant contribution of dark bursts. Thus it is unlikely that at high-z,
where metallicity and dust content is lower, the GRBs would preferentially occur in such systems. Imminent deep {\it ALMA} surveys which will be sensitive to detecting L$_{*}$ galaxies at $z\sim6$ will help constrain the dust obscured star-formation at these early cosmic times better \citep[e.g.][]{Dunlop}.

A second possibility
would imply that there may be some evolution in the efficiency of GRB production in early stellar populations (e.g., Yuksel et al. 2008, Kistler et al. 2009, Campisi et al. 2010).
Work is ongoing to determine possible biases in GRB production observationally (e.g., Schulze et al. 2015, Vergani et al. 2015, Perley et al. 2015) 
with one such analysis favoring a metallicity ceiling for GRB
production of solar, similar to the original work at low-z by \citet{Modjaz}. 
Various papers \citep{Levesque2010, Kruhler, Graham}, have shown that GRB hosts at $z<1$ preferentially seem to be low-mass, low-metallicity systems. 
The presence of hosts with super-solar metallicity in this redshift range argues against a metallicity ceiling, instead arguing in favor of a decreased efficiency for GRB production
at high metallicities.
The absence of high metallicity galaxies in the distant Universe would therefore 
increase the fraction of SFR in low metallicity galaxies and thus the increase in GRB rates relative to the SFRD could be explained by such a bias as argued by Perley et al. (2015). In other words,
GRBs are a {\it less} biased tracer of SFR at high-z than they are at low-z!

Galaxy-integrated metallicity estimates however need to be interpreted with some caution. In the Milky Way itself, there is abundant evidence for low metallicity stars in the halo. While
they are predominantly low-mass stars which do not dominate the galaxy light, a range of metallicities may be present in a galaxy even with sub-solar gas phase metallicity (or vice-versa)
making it challenging to associate a metallicity threshold/bias for GRB production. Having a larger
sample of gas-phase and absorption-line 
metallicity estimates in high-z GRB hosts compared with the distribution of metallicities in field galaxies at the same mass, both of which will only become possible with {\it JWST}
spectroscopy, will help to highlight correlations between galaxy-integrated and local-to-the-GRB metallicities and disentangle the role of metallicity as a biasing factor for GRB production \citep{Li2008}.

Clearly, the uncertainties in the star-formation rate density at $z>3$ are significant. Larger samples of GRBs
and a better understanding of the dust obscuration properties of the field galaxy sample
are crucial for assessing whether the GRB rates are consistent with the star-formation rate measured from the field galaxy sample or whether the higher GRB rate at $z>3$
is providing a window into metallicity/IMF
trends that evolve with redshift.

Looking into the future, understanding the progenitors of the high-redshift GRB population is critical for characterizing them as a robust star-formation tracer. It has recently become clear that the ionization
parameter (i.e. ionizing photon rate per baryon) in high-z star-forming galaxies is much higher than that seen in galaxies in the local Universe as inferred from the strength of nebular lines \citep{Shim2011}.
It is however unclear what the implications of the high ionization parameter for the underlying stellar population in those galaxies are. Is this a result of a top-heavy IMF? Is this due to a higher binary
fraction of stars in compact star-forming galaxies? Or is the interstellar medium in the distant galaxies so turbulent that the stars have higher angular momentum resulting in a higher production rate
of ionizing photons? While the top-heavy IMF argument can be a panacea for both the high GRB rate and the high ionization parameter, tieing the production efficiency of GRBs with the angular momentum distribution of stars is crucial for understanding whether stellar rotation may be playing a role in understanding the observational results.

A further important consideration is the potential identification of GRBs without a high-energy trigger. Now that this feat has been accomplished by the Intermediate Palomar Transient Factory (iPTF; Cenko et al. 2015), we know that with a suitable cadence such a discovery is possible from the GRB afterglow alone. The WFIRST space telescope, with a 0.3 deg$^2$ FOV, may have the ability to both discover and follow-up high-redshift GRBs from their afterglows and continue the study of new GRBs into the late 2020's beyond the time when current
high-energy GRB detection experiments are guaranteed to be operating. This in particular, motivates the development of next-generation GRB missions such as SVOM so that constraints on the X-ray and $\gamma$-ray properties of transients seen at optical/near-infrared wavelengths can be obtained. Such a multi-wavelength approach to a large sample of optical transients would help constrain
the mass range of the progenitor stars and help improve the mapping from GRB rate to the star-formation rate. 

\section{Metallicity Constraints on GRB-derived SFR}
In the previous section, we have outlined the results and uncertainties on star-formation rate densities from field galaxy surveys and from GRBs.
Here we discuss our current understanding of how observations of metallicities in damped Ly-$\alpha$ systems could constrain the star-formation rate inferred from GRBs.

We consider two star-formation rate histories that span the range that is consistent with the observational constraints presented earlier. In one, we consider 
the \citet{Behroozi} SFR which ties the stellar mass of galaxies with the corresponding dark matter halo mass and uses merger rates and mass accretion histories
to derive average star-formation histories (see also Behroozi \& Silk 2014). In another, we take the GRB rate derived SFR based on the
analysis of \citet{Kistler13} at $z>4$ folded in with the \citet{Behroozi} SFR at $z<4$.
The difference is the second model is 0.3 dex higher than the first model at high redshifts which would also be consistent with the analysis of Trenti et al. (2013). 
This latter study also
favors a metallicity origin for the increase in the GRB-to-SFR
  ratio with redshift out to $z\lesssim 4$, while at higher redshift
  their model predicts the ratio evolves primarily because the SFR
  from Lyman-break galaxy surveys is missing faint galaxies.
 
 We have used a cosmic evolution model in the context of hierarchical structure formation that follows
the interplay between the ISM and IGM due to star formation. 
This is coupled to a detailed model for cosmic chemical
evolution \citep{daigne06b,rollinde}, and allows us to 
keep track of the ionization history of the Universe, the abundances  
of the important elements produced in massive and low/intermediate mass stars as a function of redshift,
as well as the rate of core-collapse supernovae (SNII), the specific SFR (sSFR),  and the stellar mass density
\citep{vangioni}.

We show in \citet{vangioni} that using the SFR from Kistler et al. (2013),
the derived Thomson scattering optical depth is too high 
compared to the WMAP optical depth by a factor of two if we use 
the usual escape fraction value $f_{\rm esc} = 0.2$. This would imply that compared to the recent {\it Planck}-derived
value of $\tau$, the tension would be even greater. Lowering the escape fraction by a factor of two or more, addresses this issue
since it reduces the number of ionizing photons escaping the star-forming galaxy and thereby reduces the number
density of electrons that scatter the CMB photons.
This result is similar to that seen by \citet{Robertson2012} and discussed earlier.

%Note that any further flattening of the SFR at high redshift would produce a greater abundance of metals and lead to an
%even more excessive optical depth.  However, since the escape fraction of ionizing photons
%from high-z galaxies is an unconstrained number and could range from 0.1\% to 100\%, and the fact that t
%The data given in \citet{Kistler13} however was derived using a normalization of the GRB rate to the SFR based on the 
%\citet{HB06} SFR. 
%Adopting the analysis by Trenti et al. (2013) and Behroozi \& Silk (2014) leads to 
%a lower SFR at high redshift by a factor of approximately 0.3 dex \citep{Trenti}.
%The resulting fit is shown in Fig.~\ref{sfropt}.
%By a small adjustment in the slope of the SFR at high redshift, we can 
%obtain the fit to the \citet{Behroozi} normalization of the SFR based on the GRB rate
%as shown as a black line in the left panel of Fig.~\ref{sfropt}. 
%The corresponding evolution of the optical depth is shown 
%by the black curve in the right panel of Fig.~\ref{sfropt}. We now obtain a significantly 
%better fit for the optical depth, $\tau = 0.087$ using $f_{\rm esc} = 0.2$, though the redshift of reionization remains somewhat low, 
%$z_{\rm I} = 8.62$. 

\begin{figure}[h]
\begin{tabular}{ll}
\includegraphics[scale=0.255]{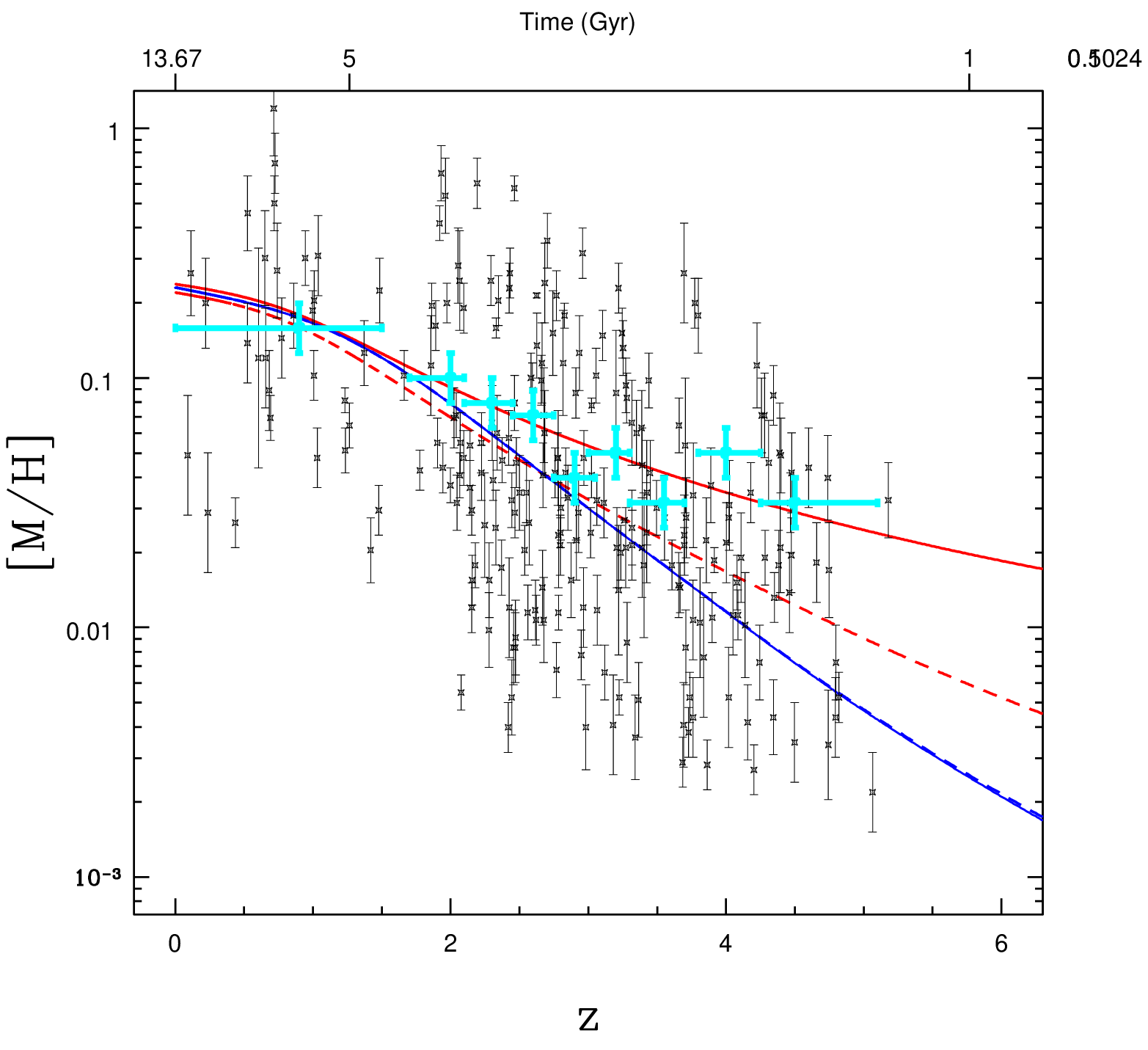}
&
\includegraphics[scale=0.255]{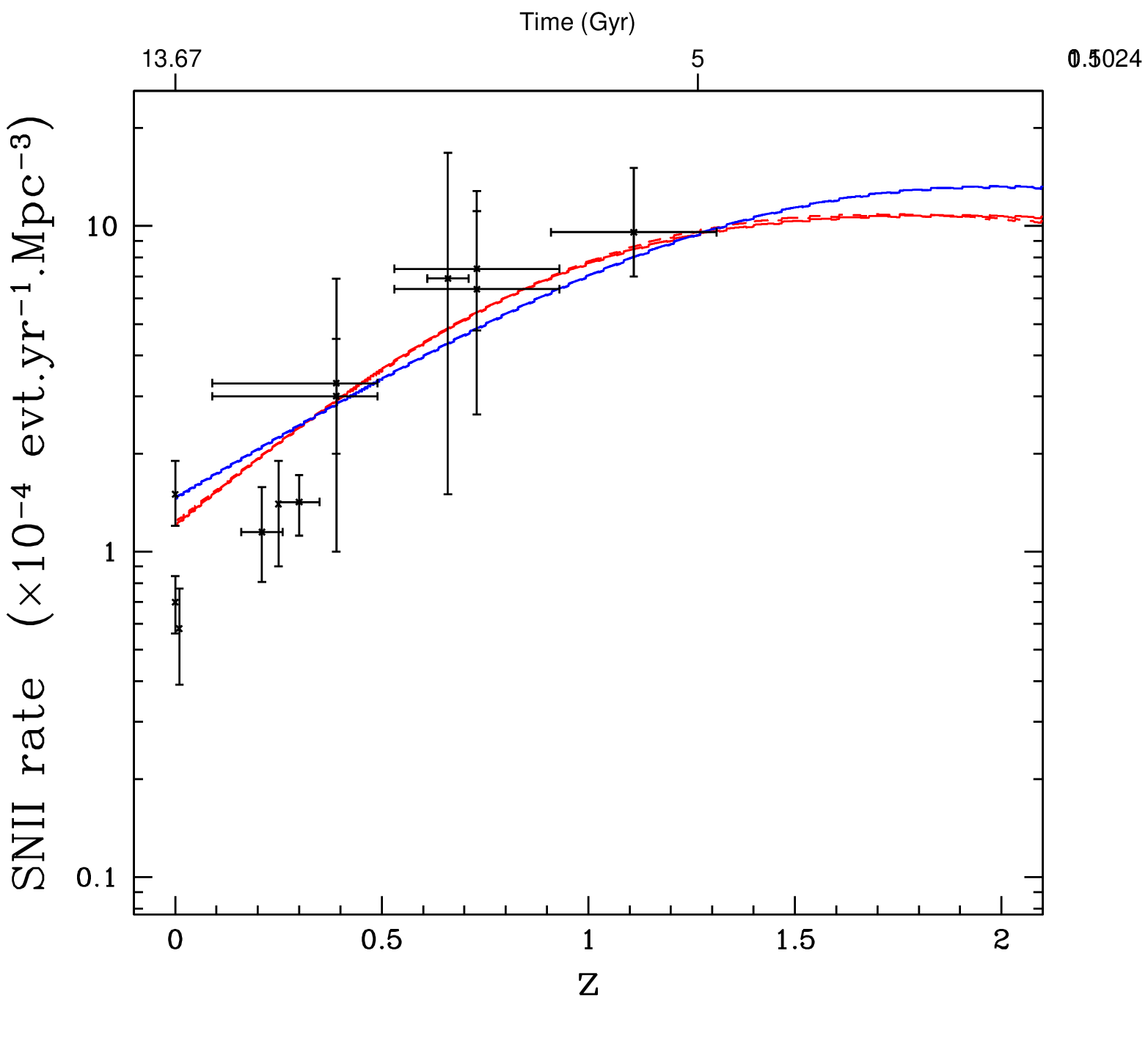}
\end{tabular}
\caption{ 
{\sl Left}: The derived (log of the) metallicity abundance relative to the solar metallicity for different
star-formation histories. The red curve corresponds to the \citet{Kistler13} SFR derived from GRBs, the blue curve
corresponds to the SFR of \citet{Behroozib} while the dashed 
red curve corresponds to the \citet{Trenti} rate. Data points are from damped Lyman-$\alpha$ systems
(Rafelski et al. 2012) with the cyan points showing the average in redshift bins. These metallicities are lower
than those plotted in Figure 3 since quasar DLAs tend to probe lower metallicity environments than the nuclei of GRB hosts which are
star-forming galaxies.
{\sl Right}: the derived core-collapse supernova rate (SNII) using the same color coding as in the left panel.
All quantities are shown as functions of redshift. See Vangioni et al. (2015) for more details. 
Clearly, the observed scatter in both metallicities and Type II SNe rates is large enough that
different star-formation histories cannot be distinguished; the preference from the metallicity measurements is 
however for a high SFR similar to that derived from GRBs.
}
\label{Metals}
\end{figure}

When instead we compare the metal yields with the metallicity of damped Ly-$\alpha$ systems reported in Rafelski et al. (2012), we find that
the two models cannot be discriminated using current observational data. Although the average metallicities are consistent
with the higher SFR of \citet{Kistler13} and \citet{Trenti}, the SNII rate which is primarily measured at $z<1.5$, is consistent with the full range of models. 
This can be seen by comparing the red solid and dashed curves in Figure~\ref{Metals} with the solid blue curve.

It can therefore be concluded that the different data sets considered as a whole favor a higher SFR at $z>3$ similar to that derived from GRBs and a low 
escape fraction (f$_{esc}<0.2$) of ionizing photons. However, the scatter in both the derived metallicities and measured $\tau$ is too large for this to be a robust conclusion.
Since the current uncertainty on
$\tau$ provides only a weak constraint on star-formation at $z>6$, we think that using the optical depth as a constraint
on star-formation is only useful once these uncertainties are improved in future analysis of {\it Planck} polarization data. 
 
\section{Metallicity Evolution of the ISM in GRB Hosts: Probing Feedback}
Spectroscopy of GRB afterglows can provide a unique window into the gas column density and metallicity of
the ISM of their hosts, the intensity of the radiation field therein and the distribution of metals along the line of sight, in some
ways the after-effects of star-formation in the galaxies.
Although these are measurements along a single line of sight through the host, by obtaining measurements of several GRBs
at similar redshifts, we can obtain an average measure of the metal abundances in sub-L$_{*}$ galaxy populations at high-z, 
a population which is challenging to characterize in any detail due to their intrinsic faintness \citep[e.g][]{Laskar}. Even with the advent of multiplexed
infrared spectrographs on 10m class telescopes, such studies on field galaxies have been limited to the bright end of the
galaxy luminosity function and typically at $z<3$ \citep{Shapley}. However, upcoming work by Faisst et al. (private communication)
which leverages the entrance of nebular lines into the {\it Spitzer}/IRAC bandpasses calibrated against rest-frame UV spectroscopy 
will help push the envelope on mass-metallicity studies out to $z\sim6$, prior to {\it JWST}. 

By comparing the redshift evolution of metallicity in GRB hosts with the corresponding evolution in stellar mass, one can
thereby characterize the nature of stellar winds, supernovae and feedback processes which may drive metals out of the host galaxy.
For example, if galaxies were a closed-box, then as long as they were star-forming, the metallicity should increase in direct proportion
to their stellar mass density. Since galaxies are not closed-box systems, the infall of pristine gas at high redshift can dilute metallicity, 
while the corresponding star-formation
fueled by that gas supply may be able to compensate for that dilution, depending on the star-formation efficiency. Furthermore,
if feedback processes are very efficient, then the gas-phase metallicity of the ISM will not increase at the same rate as the stellar mass density
since metal enrichment through star-formation is lost through galactic winds.
This effect is best seen at $z\sim0$ in the mass-metallicity relationship of galaxies seen in the Sloan Digital Sky Survey \citep{Tremonti}.
In that relationship, dwarf galaxies are five times more metal depleted that L$_{*}$ galaxies and faint galaxies show significantly larger
scatter than bright ones. This suggests that stochastic star-formation in faint galaxies combined with galactic winds efficiently remove
metals from galaxy potential wells.

\begin{figure}[h]
\center
\includegraphics[scale=0.5]{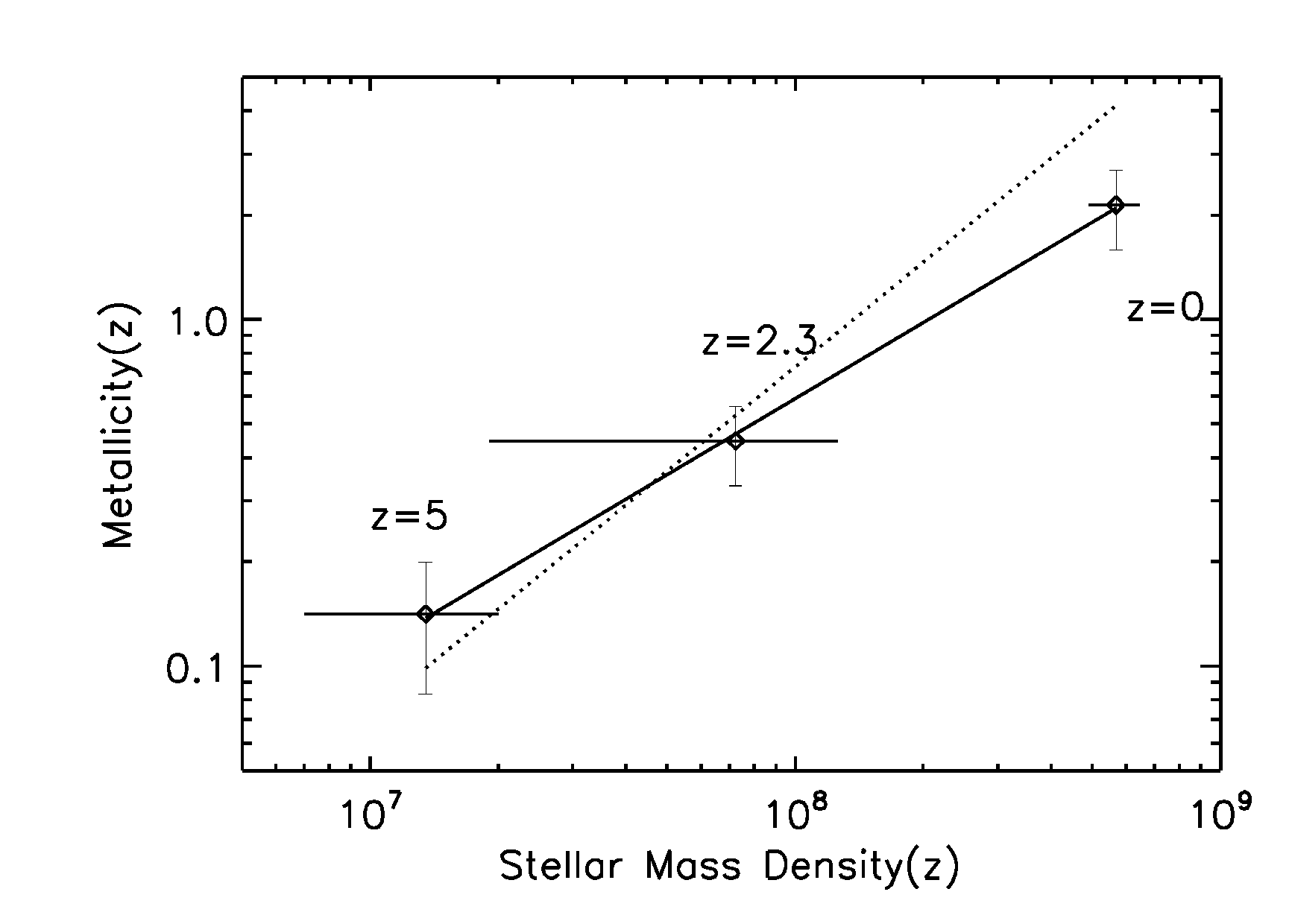}
\caption{Evolution of metallicity in GRB hosts at three different redshift bins plotted against the comoving stellar mass density (M$_{\sun}$\,Mpc$^{-3}$) at those
redshifts. The solid line is the best fit of the functional form $Z(z)\propto\rho_{*}(z)^{0.7}$ where $Z(z)$
is the metallicity as a function of redshift averaged over a multitude of GRB hosts and $\rho_{*}(z)$ is the stellar mass density dominated by faint, field galaxies. 
The dotted line has an exponent of unity. The fact that the best-fit
exponent is less than one suggests that metals are blown out of low-mass galaxies very efficiently by stellar winds, an effect which is crucial
for the difference in the faint end slope of the UV luminosity function of galaxies and the visible light luminosity function. GRBs are therefore a unique probe
of feedback in low luminosity galaxies. However, better metallicity measurements of a larger sample of high-z GRBs are required to constrain this relationship better
and to derive it as a function of halo mass.
}
\label{fig3}
\end{figure}

Figure \ref{fig3} shows the comparison between the average metallicity of GRB hosts derived from line of sight spectroscopy of the afterglow
in 3 different redshift bins compared to the stellar mass density in those bins obtained from field galaxy surveys. We find that
the best-fit relation has an exponent of 0.7 (i.e. $Z(z)\propto\rho_{*}(z)^{0.7\pm0.2}$), in the sense that metal abundances in faint galaxies increase at a slower rate than the stellar mass
density \citep{chary2007a}. If this is due to the effect of feedback processes, a similar plot generated for bright galaxies would fall closer to a line which has an exponent of unity.
Thus, through spectroscopy of the bright afterglows of GRBs, we can obtain insights into feedback processes from star-formation in faint galaxies which would
otherwise be challenging to obtain.

\section{Conclusions}
In this summary, we have highlighted the complementary role that GRBs can play to field galaxy surveys
in constraining the star-formation rate density at high-z. The GRB rate per unit star-formation derived from field galaxy surveys
appears to be increasing with increasing redshift. This may be due to a top-heavy IMF which increases the fraction of massive
stars for a particular UV luminosity density. Alternately, a metallicity bias for GRBs whereby galaxies above roughly solar metallicity
are inefficient GRB producers could explain this observational result. Since a larger fraction of star-formation at low-z takes place
in such galaxies, than at high-z, if the GRB rate per unit low-metallicity-star-formation is constant, a normalization between the GRB and total SFR
at low redshift would result in the apparent GRB rate per unit total star-formation rate increasing with increasing redshift.
A large contribution from
faint/dusty galaxies below the detection limit of field galaxy surveys could also explain the observed trend but is disfavored based on the sparsity of dust obscured
high-z galaxies in current surveys. While the Thomson scattering
optical depth of CMB photons ($\tau$) could help constrain the high-z SFR, it too is plagued by both measurement uncertainty and
the uncertainty in the escape fraction of ionizing photons from galaxies.  However, the ensemble of current data favor the high SFR (assuming a conversion from far-ultraviolet
luminosity density to SFR using a Salpeter IMF) inferred from GRB measurements,
a high ionizing photon production rate probably due to a top-heavy IMF, a low escape fraction of $\sim$10-15\% and a low clumping factor for the IGM.
Future breakthroughs in high-z star-formation will come from detecting significant populations of GRBs at $z>4-10$ with 
existing missions such as {\it Swift} and future missions such as SVOM
and targeting them through multi-wavelength spectroscopy. 
This would yield a better measurement of the luminosity and metallicity distribution
of high-z GRB hosts which can robustly discriminate between these scenarios and provide insights into galactic feedback processes at the low-mass end. 
Better measurements of high-z GRB rates and $\tau$ could
also provide a unique constraint
into the escape fraction of ionizing photons at $z\sim6$.
Finally, improved GRB progenitor models which provide greater insights on the role of
magnetic torques, angular momentum and metallicity of massive stars in the formation of GRBs will also help reduce the scatter in the mapping
of GRB rates to star-formation rates.

%\begin{figure}
%  \includegraphics[scale=0.5]{figsfr.eps}
%\caption{Fit to the SFR (versus redshift) based on the GRB rate as derived in \citet{Kistler13} using the
%normalization by \citet{Behroozi}. }
%\label{SFR}       % Give a unique label
%\end{figure}
%

%\begin{figure*}
%  \includegraphics[width=0.75\textwidth]{SFR.png}
%\end{figure*}
%

\begin{acknowledgements}
The work of EV and PPJ has been carried out at the ILP LABEX (under reference ANR-10-LABX-63) supported by French state funds managed by the ANR 
within the Investissements d'Avenir programme under reference ANR-11-IDEX-0004-02. It was also sponsored by the French Agence Nationale pour 
la Recherche (A.N.R.) via the grant VACOUL (ANR-2010-Blan-0510-01). We thank the referees for their feedback and clarifying remarks.
\end{acknowledgements}
%\bibliographystyle{aps-nameyear}      % American Physical Society (APS) style, author-year citations
%\bibliography{example}                % name your BibTeX data base
%\nocite{*}


\end{document}